\documentclass[11pt]{article} 

\usepackage[utf8]{inputenc} 

\usepackage{geometry} 
\usepackage{graphicx} 

\usepackage{booktabs} 
\usepackage{array} 
\usepackage{paralist} 
\usepackage{verbatim} 
\usepackage{subfig} 

\usepackage{fancyhdr} 
\usepackage{sectsty}
\allsectionsfont{\sffamily\mdseries\upshape} 

\usepackage[nottoc,notlof,notlot]{tocbibind} 
\usepackage[titles,subfigure]{tocloft} 

\usepackage{xcolor}

\title{Entropic Density Functional Theory}
\author{Ahmad Yousefi and Ariel Caticha \\
{\small Physics Department, University at Albany-SUNY, Albany, NY 12222, USA.}}
\date{ }
\begin{document}
\maketitle

\begin{abstract}
A formulation of the Density Functional Theory (DFT) is constructed as an
application of the method of maximum entropy for an inhomogeneous fluid in
thermal equilibrium. The use of entropy as a systematic method to generate
optimal approximations is extended from the classical to the quantum domain.
This process introduces a family of trial density operators that are
parametrized by the particle density. The optimal density operator is that
which maximizes the quantum entropy relative to the exact canonical density
operator. This approach reproduces the variational principle of DFT and
allows a simple proof of the Hohenberg-Kohn theorem at finite temperature.
Finally, as an illustration, we discuss the Kohn-Sham approximation scheme
at finite temperature.

\textbf{Keywords}: density functional theory, Hohenberg-Kohn theorem,
entropic inference, method of maximum entropy, inhomogeneous fluids.
\end{abstract}

\section{Introduction}

The Density Functional Theory (DFT) is one of the most widely used methods
for calculations of the structure of inhomogeneous many-body systems
including atoms, molecules, liquids, solids, and surfaces \cite{Kohn 1999}%
\cite{Jones 2015} (for a pedagogic introduction see \cite{Argaman Makov 2000}%
). The theory, which finds its earliest roots in the Thomas-Fermi-Dirac
model, was first introduced in its modern form by Hohenberg and Kohn who
showed that the ground state of an electron gas in an external potential can
be uniquely characterized by the electron density \cite{Hohenberg Kohn 1964}
and by Kohn and Sham who showed how to include the effects of exchange and
correlations \cite{Kohn Sham 1965}. The implications of these ideas were
soon extended to finite temperatures in the context of the grand canonical
framework \cite{Mermin 1965} (see also \cite{Eschrig 2010}-\cite{Burke et al
2016} and references therein) and to the statistical mechanics of
non-uniform classical fluids, such as the liquid-vapor interface\ \cite%
{Ebner et al 1976}-\cite{Rosenfeld 1989} (see also \cite{Evans et al 2016}
for more references).

In previous work we derived the classical DFT as an application of the
method of maximum entropy \cite{Yousefi Caticha 2021}. A central concept is
the use of entropy itself as a tool to generate optimal approximations to
probability distributions \cite{Tseng Caticha 2008} in terms of those
variables that capture the relevant physical information namely, the
particle density $n(x)$. We showed that the entropic DFT (eDFT) approach
directly leads to Evans' variational principle of the classical DFT \cite%
{Evans 1979}.

In this paper we are concerned with the very foundations of the DFT\
framework and our main goal is to extend the entropic DFT (eDFT) formalism
to the quantum domain. We emphasize that our goal is neither to derive an
alternative to DFT nor to develop improvements to the approximations that
are inevitably necessary to the successful implementation of DFT in
practical applications.
Although information theory has been used to quantify chemical concepts \cite{Alipour 2018, Rong 2020} alternative information theoretic interpretations of DFT has been suggested \cite{Nagy 2003, Nalewajski 2003} mostly based on the principle of minimum Fisher information \cite{Freiden 1998}, our reformulation stands out from three different aspects: 
First, the maximum entropy foundation on which our formulation is constructed is a completely general method of inference about quantum systems with incomplete information; regardless of the source of the information. In this interpretation of the maximum entropy, information is neither a physical quantity stored in the system, nor is an amount of uncertainty in the probability distribution or the density matrix; it is rather a constraint under which one is to update their degree of rational belief. This is important, specially, in the case of the DFT, the constraint on the density function arises from a computational assumption that the particle density function of the system is known. However this information is neither gathered by measurments nor directly obtained from the equilibrium density matrix.
Second, we work in the canonical ensemble with fixed number of particles. This liberates the foundation of DFT from the second quantization.
Third and finally, our reformulation is done for the more general DFT at finite temperature which, to our knowledge, had not been tackled by information theoretic approaches prior to this work.\\
In section 2 we review the use of relative entropy as a tool to update
density operators in response to new information and we extend the use of
entropy as a tool to derive optimal approximations from the classical
context \cite{Tseng Caticha 2008} to the quantum domain. In section 3 we
construct the entropic DFT formalism and prove a form of the Hohenberg-Kohn
theorem at finite temperature within the canonical (fixed number of
particles) framework. In section 4, as an illustration of the eDFT
formalism, we discuss the Kohn-Sham model in the local density
approximation. Finally in section 5 we summarize our conclusions.

\section{Preliminaries}

The realization that a fundamental theory such as thermodynamics should be
interpreted as an application of a general scheme for inference on the basis
of information codified into constraints can be traced to Brillouin and
Jaynes \cite{Brillouin 1952}-\cite{Jaynes 1979}. According to Jaynes -- as
motivated by the Shannon's axioms \cite{Shannon 1948} -- entropy is
interpreted as the amount of information that is missing in a probability
distribution. The preferred probability distribution is that which agrees
with what we know --- the information codified into the constraints --- but
is maximally ignorant about everything else. Thus, one is led to maximize
the entropy subject to constraints, a procedure dubbed the MaxEnt method.

A drawback of this approach is that the interpretation of entropy as an
amount of missing information is not completely satisfactory. To address
this problem Shore and Johnson \cite{Shore Johnson 1980} proposed that one
could avoid invoking questionable measures of information by directly
axiomatizing the method for updating probabilities through a variational
principle that involved maximizing an entropy functional satisfying certain
desirable properties. The question of why should one adopt a variational
principle was later\textbf{\ }clarified by Skilling \cite{Skilling 1988} who
proposed a simple ranking strategy: in order to select an optimal
distribution (he had in mind the more general case of positive additive
distributions which include \emph{e.g.} intensities in an image) one
proceeds by ranking the distributions according to some preference criteria
and then choosing the one which ranks the highest. The ranking scheme is
naturally implemented by associating a real number -- the entropy -- to each
distribution with the preference criteria fixed through the axioms of Shore
and Johnson. In later work the nature of the method of maximum entropy was
further streamlined as a scheme \emph{designed} to update probabilities when
confronted with new information. In this approach the question
\textquotedblleft what is information?\textquotedblright\ receives a very
simple answer. \emph{Information is just the constraints we decide to impose
on our beliefs,} and there is no need to define \textquotedblleft
amounts\textquotedblright\ of information. The motivation behind the design
criteria was clarified and their number reduced from five to two \cite%
{Caticha 2004}-\cite{Vanslette 2017} (reviewed in \cite{Caticha 2023} and 
\cite{Caticha 2021a}).

\subsection{The quantum MaxEnt method}

The task of extending the method of maximum entropy to the quantum domain as
a method to update density operators was carried out by Vanslette \cite%
{Vanslette 2017}. The goal is to update a prior density operator $\hat{\sigma%
}$ when provided with new information in the form of the expected value of
some self-adjoint operators $\langle \hat{A}_{i}\rangle =A_{i}$. Vanslette
showed that the Umegaki relative entropy \cite{Umegaki 1962}, 
\begin{equation}
S_{r}[\hat{\rho}|\hat{\sigma}] = {\textrm{Tr}} \left( \hat{\rho}\log \hat{\rho}-\hat{\rho}\log \hat{\sigma}\right) ~,  \label{Sr} 
\end{equation}%
provides the \emph{unique} criterion to rank density operators $\hat{\rho}$
relative to the prior $\hat{\sigma}$.

The maximization of $S_{r}[\hat{\rho}|\hat{\sigma}]$ subject to the
constraints $\langle \hat{A}_{i}\rangle =A_{i}$ and normalization, 
\begin{equation}
\delta \left[ S_{r}[\hat{\rho}|\hat{\sigma}]+\alpha _{0}(1- {\textrm{Tr}}\hat{%
\rho})+\sum\nolimits_{i}\alpha _{i}(A_{i}- {\textrm{Tr}}\hat{\rho}\hat{A}_{i})%
\right] =0~,
\end{equation}%
leads to the posterior density operator 
\begin{equation}
\hat{\rho}^{\ast }=\frac{1}{Z}\exp \left( \log \hat{\sigma}%
-\sum\nolimits_{i}\alpha _{i}\hat{A}_{i}\right) ~,  \label{3}
\end{equation}%
where 
\begin{equation}
Z(\alpha _{i})=e^{\alpha _{0}}= {\textrm{Tr}}\exp \left( \log \hat{\sigma}%
-\sum\nolimits_{i}\alpha _{i}\hat{A}_{i}\right) ~.  \label{3a}
\end{equation}%
Substituting $\hat{\rho}^{\ast }$\ back into eq.(\ref{Sr}) gives the value
of the maximized entropy, 
\begin{equation}
S(A_{i}) = S_{r}[\hat{\rho}^{\ast }|\hat{\sigma}%
]=\sum\nolimits_{i}\alpha _{i}A_{i}+\log Z~.  \label{4}
\end{equation}

It is widely known that the classical MaxEnt method leads to a mathematical
formalism characterized by a contact structure (see \emph{e.g.}, \cite%
{Rajeev 2008}\cite{Balian Valentin 2001}). In a parallel development the use
of Legendre transforms in the context of DFT has also been widely explored 
\cite{Eschrig 2010}\cite{Lieb 1983}\cite{Fukuda et al 1994}. These results
can be extended to the quantum domain leading to a similar contact
structure\ (see \emph{e.g.}, \cite{Yousefi thesis 2021}). The significance
of these results is that the physical content of the formalism is preserved
under Legendre transformations quite independently of restrictions to
thermal equilibrium and of the physical significance of the so-called
\textquotedblleft free energies\textquotedblright\ or Massieu functions.

\subsection{Optimal approximations of density operators}

The last prerequisite for the construction of the DFT formalism is a
systematic method of approximation for density operators. The method we
adopt is an extension of the technique developed by Tseng and Caticha in the
classical context \cite{Tseng Caticha 2008}. The problem is that the exact
probability distributions $Q$ obtained using the MaxEnt method are often too
intractable to be useful in actual calculations. The solution is to consider
a family of more tractable trial distributions $P_{\theta }$ dependent on
some parameters $\theta $. The goal is to select the trial distribution $%
P_{\theta ^{\ast }}$ that best approximates the exact distribution $Q$. In 
\cite{Tseng Caticha 2008} it was argued that the criterion to select the
optimal parameters $\theta ^{\ast }$ is again provided by the method of
maximum entropy: The optimal $P_{\theta ^{\ast }}$ is that which is
\textquotedblleft closest\textquotedblright\ to the exact $Q$ in the sense
that it maximizes the relative entropy $S[P_{\theta }|Q]$.

Next, we extend this approximation technique to the quantum domain. We
consider a family of tractable density operators $\hat{\rho}_{\theta }$
parametrized by parameters $\theta $. The member of the trial family $\hat{%
\rho}_{\theta }$ that best approximates the exact density operator $\hat{\rho%
}^{\ast }$ is the one which maximizes the entropy of $\hat{\rho}_{\theta }$
relative to $\hat{\rho}^{\ast }$, 
\begin{equation}
\left. \frac{\partial }{\partial \theta }S_{r}[\hat{\rho}_{\theta }|\hat{\rho%
}^{\ast }]\right\vert _{\theta =\theta ^{\ast }}=0~.  \label{A10}
\end{equation}%
As an example, consider the special case where $\hat{\rho}^{\ast }$ and $%
\hat{\rho}_{\theta }$ take the exponential form, 
\begin{equation}
\hat{\rho}^{\ast }=\frac{1}{Z}e^{-\beta \hat{H}}\quad \textrm{and}\quad \hat{%
\rho}_{\theta }=\frac{1}{Z_{\theta }}e^{-\beta \hat{H}_{\theta }}~,
\label{A11}
\end{equation}%
where $\hat H$ and $\hat H_\theta$'s are some Hermitian operators of interest,
the Gibbs inequality, 
\begin{equation}
S_{r}[\hat{\rho}_{\theta }|\hat{\rho}^{\ast }]\leq 0~,
\end{equation}%
reduces to the Bogolyubov inequality, 
\begin{equation}
F\leq F_{\theta }+\langle \hat{H}-\hat{H}_{\theta }\rangle _{\theta }~,
\label{A12}
\end{equation}%
where 
\begin{equation}
F=-\frac{1}{\beta }\log Z~,\quad F_{\theta }=-\frac{1}{\beta }\log Z_{\theta
}~,\quad \textrm{and}\quad \langle \cdot \rangle _{\theta }={\textrm{Tr}}\left[ 
\hat{\rho}_{\theta }(\cdot )\right] ~.  \label{A13}
\end{equation}%
Thus, the argument above shows the popular approximation method based on the
Bogolyubov inequality (see \emph{e.g.}, \cite{Feynman 1972}) is a special
case of the more general approximation method based on entropy maximization.

\section{Density functional formalism}

The goal of the DFT formalism is to find tractable approximations to study
the structure of matter. The first crucial step is to recognize that the
quantity that captures the desired structural information is the electron
density $n(x)$. We wish to design a formalism in which the central role
played by the electron density is explicitly displayed.

In the absence of magnetic fields the time independent Schr\"{o}dinger
equation for an electron gas of $N$ particles is 
\begin{equation}
\hat{H}|\psi \rangle =E|\psi \rangle ~,  \label{5}
\end{equation}%
where 
\begin{equation}
\hat{H}_{v}=\hat{H}^{(0)}+\hat{V}=\hat{K}+\hat{U}+\hat{V}=\sum%
\limits_{i=1}^{N}\frac{\hat{p}_{i}^{2}}{2m}+\frac{e^{2}}{2}%
\sum\limits_{j\neq k}^{N}\frac{1}{|\hat{x}_{j}-\hat{x}_{k}|}%
+\sum\limits_{l=1}^{N}v(\hat{x}_{l})~,  \label{6}
\end{equation}%
and $|\psi \rangle $ is an antisymmetrized product of $N$ two-spinor
orbitals. The potential $\hat{U}$ describes interparticle interactions and
the potential $\hat{V}$ \ describes interactions with nuclei and other
external potentials.

\subsection{Introducing density as the relevant variable}

We are interested in the thermal properties of an inhomogeneous electron
fluid and therefore we need trial states that describe both thermal
equilibrium and inhomogeneity. The former is imposed by a constraint on the
expected value of energy and the latter is incorporated by constraints on
the expected value $n(x)$ of the electron density $\hat{n}(x)$. The
continuous density function $n(x)$ plays a role analogous to the discrete
parameters $\theta $ in equations (\ref{A10}-\ref{A13}).

Adopting a uniform prior, the relevant trial states are obtained by
maximizing the entropy 
\begin{equation}
S_{r}[\hat{\rho}|\hat{1}]=-\textrm{Tr}\hat{\rho}\log \hat{\rho}~,  \label{7}
\end{equation}%
subject to the constraints 
\begin{eqnarray}
{\textrm{Tr}}\hat{\rho} &=&1~,  \label{8} \\
{\textrm{Tr}}\hat{\rho}\hat{H}_{v} &=&E~,  \label{9} \\
\textrm{and\quad }{\textrm{Tr}}\hat{\rho}\hat{n}(x) &=&n(x)~,  \label{10}
\end{eqnarray}%
where 
\begin{equation}
\hat{n}(x)=\sum\limits_{i=1}^{N}\delta \left( \hat{x}_{i}-x\right) \quad 
\textrm{and} \quad \int d^{3}x\,n(x)=N~.  \label{10a}
\end{equation}%
To be clear, throughout this work the trace is taken over the Hilbert space
of a fixed number $N$ of particles and in this respect our formalism \emph{%
resembles} the canonical ensemble approach. Indeed, all states $|\psi
\rangle $ in the Hilbert space are eigenstates of the number operator, 
\begin{equation}
\hat{N}|\psi \rangle =\int d^{3}x\,\hat{n}(x)|\psi \rangle =N|\psi \rangle
\quad \textrm{so that}\quad \langle \psi |\hat{N}|\psi \rangle =N~,
\label{10b}
\end{equation}%
but they need not be eigenstates of the density operators $\hat{n}(x)$. Our
formalism \emph{differs} from the canonical formalism in that eq.(\ref{10})
represents an additional infinite number of constraints --- one constraint
on the expected density function $n(x)$ at each point in space. Due to (\ref%
{10b}) the expected density function $n(x)$ is not arbitrary; it is
constrained to obey (\ref{10b}).

Proceeding to the MaxEnt analog of eq.(\ref{3}) we find the trial density
operator 
\begin{equation}
\hat{\rho}_{n}=\frac{1}{Z_{v}}\exp \left( -\beta \hat{H}_{v}-\int
d^{3}x\,\alpha (x)\hat{n}(x)\right) ~,  \label{11}
\end{equation}%
where 
\begin{equation}
Z_{v}(\beta ;\alpha ]={\textrm{Tr}}\exp \left( -\beta \hat{H}_{v}-\int
d^{3}x\,\alpha (x)\hat{n}(x)\right) ~,  \label{12}
\end{equation}%
and where $\beta $ and the infinite number of Lagrange multipliers $\alpha
(x)$ are implicitly determined by 
\begin{equation}
\frac{\partial \log Z_{v}(\beta ;\alpha ]}{\partial \beta }=-E\quad \textrm{and%
}\quad \frac{\delta \log Z_{v}(\beta ;\alpha ]}{\delta \alpha (x)}=-n(x)~,
\label{12a}
\end{equation}%
with the additional constraint (\ref{10a}), 
\begin{equation}
\int d^{3}x\,n(x)=-\int d^{3}x\frac{\delta Z_{v}(\beta ;\alpha ]}{\delta
\alpha (x)}=N~.  \label{14}
\end{equation}

The notation $Z_{v}(\beta ;\alpha ]$ indicates that $Z$ is a function of $%
\beta $ and a functional of $\alpha (x)$ and depends on $v(x)$ through the
Hamiltonian $\hat{H}_{v}$. At this point in the argument there is no
implication that the trial states $\hat{\rho}_{n}$ are in any way more
computationally tractable than the exact state $\hat{\rho}^{\ast }$ obtained
from (\ref{11}) by setting $\alpha (x)$ to zero.

Next we calculate the entropy of $\hat{\rho}_{n}$ relative to the uniform
prior to define the trial entropy, 
\begin{equation}
S_{r}[\hat{\rho}_{n}|\hat{1}]=\beta E+\int d^{3}x\,\alpha (x)n(x)+\log
Z_{v}(\beta ;\alpha ]\ {=}S_{v}(E;n]~.  \label{13}
\end{equation}

An important symmetry of the DFT formalism, which is what makes the whole
DFT\ formalism work, arises from the fact that the dependence of $\hat{\rho}%
_{n}$ and $Z_{v}(\beta ;\alpha ]$ on $v(x)$ and $\alpha (x)$ occurs only
through the particular combination 
\begin{equation}
\alpha _{{int}}(x){=}\alpha (x)+\beta v(x)~.
\label{15}
\end{equation}%
The reason for the subscript `int', which denotes `intrinsic', will become
clear later in eq.(\ref{mu int a}). This DFT symmetry implies that a change
in the potential $v(x)$ can be compensated by a suitable change in the
multiplier $\alpha (x)$ in such a way that $\alpha _{{int}}(x)$ and the
expected density $n(x)$ remain unaffected. From (\ref{6}) and (\ref{15}) we
find that (\ref{12}) can be written as 
\begin{equation}
Z_{v}(\beta ;\alpha ]={\textrm{Tr}}\exp \left( -\beta \hat{H}^{(0)}-\int
d^{3}x\,\alpha _{{int}}(x)\hat{n}(x)\right) {=}%
Z(\beta ;\alpha _{{int}}]~,  \label{16}
\end{equation}%
so that eqs.(\ref{11}) and (\ref{12a}) become 
\begin{equation}
\hat{\rho}_{n}=\frac{1}{Z(\beta ;\alpha _{{int}}]}\exp \left( -\beta 
\hat{H}^{(0)}-\int d^{3}x\,\alpha _{{int}}(x)\hat{n}(x)\right)
\label{17}
\end{equation}%
and 
\begin{equation}
\quad n(x)=-\frac{\delta \log Z(\beta ;\alpha _{int}]}{\delta \alpha_{int}(x)} ~.  \label{17b}
\end{equation}

\subsection{ The entropic DFT variational principle}

The exact canonical density operator $\hat{\rho}^{\ast }$ is found by
maximizing (\ref{7}) subject to (\ref{8}) and (\ref{9}). The result can be
read off eq.(\ref{11}) by setting $\alpha (x)=0$, 
\begin{equation}
\hat{\rho}^{\ast }=\frac{1}{Z_{v}(\beta )}\exp \left( -\beta \hat{H}%
_{v}\right) \quad \textrm{and}\quad Z_{v}(\beta )= {\textrm{Tr}}\exp \left(
-\beta \hat{H}_{v}\right) ~.  \label{18}
\end{equation}%
The goal is to
approximate $\hat{\rho}^{\ast }$ by the best matching member of the family $%
\{\hat{\rho}_{n}\}$ with all density operators referring to the same $\beta $
and $N$. This involves maximizing the entropy of $\hat{\rho}_{n}$ relative
to $\hat{\rho}^{\ast }$, 
\begin{equation}
\left. \frac{\delta S_{r}[\hat{\rho}_{n}|\hat{\rho}^{\ast }]}{\delta n(x)}%
\right\vert _{\beta ,N}=0~.  \label{20}
\end{equation}%
From (\ref{11}) and (\ref{18}) we find 
\begin{equation}
S_{r}[\hat{\rho}_{n}|\hat{\rho}^{\ast }]=\int d^{3}x\,\alpha (x)n(x)+\log
Z_{v}(\beta ;\alpha ]-\log Z_{v}(\beta )~.  \label{19}
\end{equation}

Introducing a Lagrange multiplier $\alpha ^{\ast }$ to enforce the
constraint on $N$ we have, 
\begin{equation}
\frac{\delta }{\delta n(x)}\left[ S_{r}[\hat{\rho}_{n}|\hat{\rho}^{\ast
}]+\alpha ^{\ast }\left( N-\int d^{3}x^{\prime }\,n(x^{\prime })\right) %
\right] _{\beta }=0~.  \label{eDFT a}
\end{equation}%
From the construction above one might expect that the optimal $\hat{\rho}%
_{n} $ coincides with the exact $\hat{\rho}^{\ast }$. We can check that this
is indeed the case. Substituting eq.(\ref{19}) into (\ref{eDFT a}) we find 
\begin{equation}
\int d^{3}x^{\prime }\,\left[ n(x^{\prime })+\frac{\delta \log Z_{v}(\beta
;\alpha ]}{\delta \alpha (x^{\prime })}\right] \frac{\delta \alpha
(x^{\prime })}{\delta n(x)}=\alpha ^{\ast }-\alpha (x)  \label{23}
\end{equation}%
The LHS vanishes by eq.(\ref{12a}). Therefore, the optimal $\hat{\rho}_{n}$
is achieved for $\alpha (x)=\alpha ^{\ast }$. From (\ref{11}), (\ref{18})
and (\ref{19}) we see that $\alpha ^{\ast }=0$ which means that imposing the 
$N$ constraint was unnecessary: the optimal density reproduces the exact
density $n^{\ast }(x)$ whether the variations $\delta n(x)$ preserve the
total $N$ or not.

We conclude that the entropic DFT variational principle, 
\begin{equation}
\left. \frac{\delta S_{r}[\hat{\rho}_{n}|\hat{\rho}^{\ast }]}{\delta n(x)}%
\right\vert _{n^{\ast }(x)}=0~,  \label{eDFT b}
\end{equation}%
leads to an optimal $\hat{\rho}_{n}$ which coincides with the exact
canonical $\hat{\rho}^{\ast }$ in eq.(\ref{18}), 
\begin{equation}
\hat{\rho}_{n}^{{opt}}=\hat{\rho}^{\ast },\quad \textrm{where}\quad
\alpha ^{{opt}}(x)=\alpha ^{\ast }=0~,  \label{24}
\end{equation}%
Thus, at this point our \textquotedblleft approximation\textquotedblright\
scheme is (trivially) exact: by explicit construction we have demonstrated
the existence of a functional of the density $n(x)$, $\beta $ and $N$ ---
the relative entropy $S_{r}[\hat{\rho}_{n}|\hat{\rho}^{\ast }]$ --- that
assumes its maximum value at the exact density $n^{\ast }(x)$. At this
point, however, we have not yet shown that this variational principle is
equivalent to the thermal DFT principle derived by Mermin \cite{Mermin 1965}%
. This, we show next.

\subsection{The DFT theorem}

Equations (\ref{13}) and (\ref{19}) allows us to write 
\begin{equation}
S_{r}[\hat{\rho}_{n}|\hat{\rho}^{\ast }]=-\beta \Omega _{v}(\beta ;n]-\log
Z_{v}(\beta )
\end{equation}%
where we have introduced the \textquotedblleft free
energy\textquotedblright\ functional%
\begin{equation}
\Omega _{v}(\beta ;n] {=}E-\frac{1}{\beta }S_{v}(E;n]\ .
\label{27}
\end{equation}%
The new functional $\Omega _{v}$, 
\begin{equation}
\Omega _{v}(\beta ;n]=-\frac{1}{\beta }\int d^{3}x\,\alpha (x)n(x)-\frac{1}{%
\beta }\log Z_{v}(\beta ;\alpha ]~,  \label{27a}
\end{equation}%
allows us to rewrite the entropic variational principle (\ref{eDFT a}) as 
\begin{equation}
\left. \frac{\delta \Omega _{v}(\beta ;n]}{\delta n(x)}\right\vert _{n^{\ast
}(x)}=0~.  \label{eDFT c}
\end{equation}%
The optimal density $n^{\ast }(x)$ is found by minimizing $\Omega _{v}(\beta
;n]$ at fixed $\beta $ and $N$. Furthermore, from (\ref{27a}) the
multipliers $\alpha (x)$ are obtained from 
\begin{equation}
\alpha (x)=-\beta \frac{\delta \Omega _{v}(\beta ;n]}{\delta n(x)}~.
\label{28a}
\end{equation}%
From eq.(\ref{24}), $\alpha ^{{opt}}(x)=\alpha ^{\ast }= {const}$%
, we obtain 
\begin{equation}
\left. \nabla \frac{\delta \Omega _{v}(\beta ;n]}{\delta n(x)}\right\vert
_{n^{\ast }(x)}=0~,  \label{29}
\end{equation}%
which has been called the \textquotedblleft core integro-differential
equation of DFT\textquotedblright\ \cite{Evans 1979}.

To proceed further, substitute (\ref{6}), (\ref{9}), into (\ref{27}) to find 
\begin{equation}
\Omega _{v}(\beta ;n]=\langle \hat{K}+\hat{U}\rangle _{\hat{\rho}_{n}}+\int
d^{3}x\,v(x)n(x)-\frac{1}{\beta }S_{v}(E;n]~,  \label{32}
\end{equation}%
so that 
\begin{equation}
\Omega _{v}(\beta ;n]=F_{v}(\beta ;n]+\int d^{3}x\,v(x)n(x)~,  \label{34}
\end{equation}%
where we have introduced 
\begin{equation}
F_{v}(\beta ;n] {=}\langle \hat{K}+\hat{U}\rangle _{\hat{%
\rho}_{n}}-\frac{1}{\beta }S_{v}(E;n]~.  \label{33}
\end{equation}%

\noindent \textbf{The Density Functional Theorem:} \emph{The density
functional }$F_{v}[n]$\emph{\ is independent of the external potential }$%
v(x) $\emph{,} 
\begin{equation}
\left. \frac{\delta F_{v}(\beta ;n]}{\delta v(x)}\right\vert _{\beta
,n(x)}=0~.  \label{35}
\end{equation}%
This result justifies dropping the index $v$, 
\begin{equation}
F(\beta ;n]{=}F_{v}(\beta ;n]~,  \label{40}
\end{equation}%
and referring to $F(\beta ;n]$ as the \emph{intrinsic density functional}.
(The term `intrinsic' indicates that $F(\beta ;n]$ is independent of the
external potential $v(x)$.)

\noindent \textbf{Proof:} The crucial observation behind the entropic DFT
formalism is that $\hat{\rho}_{n}$ and $Z_{v}(\beta ;\alpha ]$ depend on the
external potential $v(x)$ and the Lagrange multiplier function $\alpha (x)$
only through the particular combination $\alpha _{{int}}(x)$ defined in
(\ref{15}). Substitute (\ref{13}), (\ref{15}) and (\ref{16}) into (\ref{33})
to get 
\begin{equation}
F_{v}(\beta ;n]=-\frac{1}{\beta }\int d^{3}x\,\alpha _{{int}}(x)n(x)-%
\frac{1}{\beta }\log Z(\beta ;\alpha _{{int}}]~.  \label{36}
\end{equation}%
Then the derivative $\delta /\delta v(x^{\prime })$ at fixed $\beta $ and $%
n(x)$ is 
\begin{equation}
\frac{\delta F_{v}(\beta ;n]}{\delta v(x^{\prime })}=\int d^{3}x^{\prime
\prime }\frac{\delta F_{v}(\beta ;n]}{\delta \alpha _{{int}}(x^{\prime
\prime })}\left. \frac{\delta \alpha _{{int}}(x^{\prime \prime })}{%
\delta v(x^{\prime })}\right\vert _{\beta ,n(x)}.  \label{37}
\end{equation}%
Eq.(\ref{17}) shows that keeping $n(x)$ fixed is achieved by keeping $\alpha
_{{int}}(x)$ fixed and vice versa, therefore%
\begin{equation}
\left. \frac{\delta \alpha _{{int}}(x^{\prime \prime })}{\delta
v(x^{\prime })}\right\vert _{\beta ,n(x)}=\left. \frac{\delta \alpha _{{%
int}}(x^{\prime \prime })}{\delta v(x^{\prime })}\right\vert _{\beta ,\alpha
_{{int}}(x)}=0~,  \label{37a}
\end{equation}%
which implies (\ref{35}) and concludes the proof.

Equations (\ref{28a}) and (\ref{29}) suggest that (up to an additive
constant) the multiplier $\alpha (x)$ plays a role \emph{analogous} to that
of a chemical potential. Let us then use eq.(\ref{28a}) to introduce 
\begin{equation}
\gamma (x){=}-\frac{\alpha (x)}{\beta }=\frac{\delta
\Omega _{v}(\beta ;n]}{\delta n(x)}~,  \label{41}
\end{equation}%
which we shall call the \textquotedblleft \emph{local} \emph{chemical
potential.}\textquotedblright\ The core eq. (\ref{29}) has a natural
interpretation: the condition for neighboring volume elements to be in
equilibrium is that the local chemical potential be uniform, 
\begin{equation}
\left. \nabla \gamma (x)\right\vert _{n^{\ast }}=0~.  \label{41a}
\end{equation}%
The optimal value of $\gamma (x)$ is 
\begin{equation}
\gamma ^{\ast }=-\frac{\alpha ^{\ast }}{\beta }=0\quad \textrm{so that}\quad
\nabla \gamma ^{\ast }=0~.  \label{41b}
\end{equation}

From eq.(\ref{34}) we have 
\begin{equation}
\delta \Omega _{v}(\beta ;n]=\delta F(\beta ;n]+\int d^{3}x\left[
n(x)\delta v(x)+v(x)\delta n(x)\right] ~,
\end{equation}%
while eq.(\ref{41}) gives 
\begin{equation}
\delta \Omega _{v}(\beta ;n]=\int d^{3}x\left( \frac{\delta \Omega _{v}}{%
\delta v(x)}\delta v(x)+\frac{\delta \Omega _{v}}{\delta n(x)}\delta
n(x)\right) =\int d^{3}x\left[ n(x)\delta v(x)+\gamma (x)\delta n(x)\right]
~.
\end{equation}%
Subtracting these two equations gives 
\begin{equation}
\delta F(\beta ;n]=\int d^{3}x\left[ \gamma (x)-v(x)\right] \delta n(x)~,
\end{equation}%
which shows that $\delta F/\delta n$ can be interpreted as the \emph{local} 
\emph{intrinsic} \emph{chemical potential}, 
\begin{equation}
\frac{\delta F(\beta ;n]}{\delta n(x)} = \gamma _{{%
int}}(x)~,  \label{gamma int a}
\end{equation}%
with 
\begin{equation}
\gamma (x)=\gamma _{{int}}(x)+v(x)\quad \textrm{and}\quad \gamma _{{%
int}}(x)=-\frac{\alpha _{{int}}(x)}{\beta }~.  \label{mu int a}
\end{equation}%
Evaluating at $n^{\ast }$ gives the \emph{equilibrium} intrinsic chemical
potential, 
\begin{equation}
\left. \frac{\delta F(\beta ;n]}{\delta n(x)}\right\vert _{n^{\ast }}=\gamma
_{{int}}^{\ast }(x)~.  \label{mu*int b}
\end{equation}%
(The term `intrinsic' reminds us that both $\gamma _{{int}}(x)$ and $%
\gamma _{{int}}^{\ast }(x)$ are independent of the external potential $%
v(x)$.)

We mentioned earlier that the multiplier $\alpha (x)$ plays a role \emph{%
analogous} to that of a chemical potential. We can now be more explicit. Let 
\begin{equation}
\mu (x)=\mu (x;n]\quad \textrm{and}\quad \mu _{{int}}(x)=\mu _{{int}%
}(x;n]
\end{equation}%
be the actual chemical potential and the intrinsic chemical potential at
location $x$, 
\begin{equation}
\mu (x;n]=\mu _{{int}}(x;n]+v(x)~.
\end{equation}%
Equilibrium among different volume elements is achieved when%
\begin{equation}
\left. \nabla \mu (x;n]\right\vert _{n^{\ast }}=0\Longrightarrow \mu
(x;n^{\ast }]=\mu ^{\ast }= {const}.
\end{equation}%
Then eqs.(\ref{41})-(\ref{41b}) lead us to identify 
\begin{equation}
\gamma (x)=\mu (x)-\mu ^{\ast }\quad \textrm{so that}\quad \gamma ^{\ast }=0~,
\end{equation}%
and 
\begin{equation}
\gamma _{{int}}(x)=\mu _{{int}}(x;n]-\mu (x;n^{\ast }]~.
\label{gamma int b}
\end{equation}

We can express the eDFT variational principle in terms of $F$. Using eqs.(%
\ref{eDFT c}) and (\ref{34}) we find 
\begin{equation}
\left. \frac{\delta }{\delta n(x)}\left( F(\beta ;n]+\int d^{3}x^{\prime
}\,v(x^{\prime })n(x^{\prime })\right) \right\vert _{n^{\ast }(x)}=0~.
\label{eDFT d}
\end{equation}%
To summarize, we have reproduced the foundational theorem behind the thermal
DFT\ formalism as an application of maximum entropy methods. This is the
main result of this paper. The treatment, so far, has been exact. In the
next section, as an illustration of the method, we adapt the well-known
Kohn-Sham model to the entropic DFT approach.

\section{The Kohn-Sham approximation scheme}

\label{KS scheme} The exact calculation of $F(\beta ;n]$ requires
calculating $Z(\beta ;\alpha _{{int}}]$. Unfortunately, this is just as
difficult as calculating the original canonical partition function $%
Z_{v}(\beta )$ which was precisely what we wanted to avoid. An analogous
problem arises in the standard many-body theory: even for relatively small
particle numbers the calculation of the $N$-particle wave function becomes
impractically difficult because the wave function $\Psi (\vec{r}_{1}\ldots 
\vec{r}_{N})$ lives in a $3N$-dimensional configuration space. The DFT
framework attempts to evade this problem by focusing attention on the
hopefully easier problem of calculating the density $n(x)$ which is a \emph{%
function} that lives in a mere $3$ dimensions. Unfortunately, the problem is
not solved, but merely transferred to the calculation of the \emph{functional%
} $F(\beta ;n]$. Not all is lost, however, because the reformulation in
terms of the density $n(x)$ suggests new useful approximations.

The discussion below parallels closely the ground state formulation of Kohn
and Sham \cite{Kohn Sham 1965}. It differs from the grand canonical thermal
DFT of Mermin \cite{Mermin 1965} in that here we remain within the canonical
framework of fixed particle number. In common with the Hartree-Fock
approximation the Kohn-Sham model reduces an interacting many-particle Schr%
\"{o}dinger equation to that of a single particle in the presence of an
effective potential that includes exchange and correlation effects. An
important advantage is that, unlike Hartree-Fock, the Kohn-Sham framework
can \emph{in principle} be exact. In practice, however, the success of the
model hinges on whether the approximations for exchange and correlations are
sufficiently simple and accurate. Fortunately, the \textquotedblleft local
density approximation,\textquotedblright\ which is exact for a uniform
electron gas, and should remain valid for slowly varying potentials, has
turned out to be quite successful for the prediction of bond lengths and
molecular structures even when these involve inhomogeneities at the atomic
scale.

Referring to eq.(\ref{33}) the idea is that $F(\beta ;n]$ can be split into
three terms, 
\begin{equation}
F(\beta ;n]=F_{0}(\beta ;n]+U_{C}[n]+F_{\textrm{xc}}(\beta ;n]~.  \label{KS a}
\end{equation}%
The first term $F_{0}(\beta ;n]$ represents the intrinsic free energy of a
gas of \emph{non-interacting} and \emph{uncorrelated} particles at the same
temperature and density. The second term $U_{C}[n]$ is the classical Coulomb
interaction, 
\begin{equation}
U_{C}[n]=\frac{e^{2}}{2}\int d^{3}xd^{3}x^{\prime }\frac{n(x)n(x^{\prime })}{%
|x-x^{\prime }|}~,
\end{equation}%
that represents the dominant contribution from the interparticle potential
term $\langle \hat{U}\rangle _{\hat{\rho}_{n}}$ in (\ref{33}). The third $F_{%
\textrm{xc}}(\beta ;n]$ is a correction that accounts for all additional
exchange and correlations effects. To the extent that we can define $F_{%
\textrm{xc}}(\beta ;n]$ to be the difference%
\begin{equation}
F_{\textrm{xc}}(\beta ;n]{=}F(\beta ;n]-F_{0}(\beta
;n]-U_{C}[n]~,
\end{equation}%
equation (\ref{KS a}) is trivially exact.

We are now ready to substitute (\ref{KS a}) into the eDFT variational
principle (\ref{eDFT d}). The result is 
\begin{equation}
\left[ \frac{\delta F_{0}}{\delta n(x)}+v(x)+\int d^{3}x^{\prime }\frac{%
e^{2}n(x^{\prime })}{|x-x^{\prime }|}+v_{\textrm{xc}}(x;n]\right] _{n^{\ast
}(x)}=0~,  \label{eDFT e}
\end{equation}%
where we introduced 
\begin{equation}
v_{\textrm{xc}}(x;n]{=}\frac{\delta F_{\textrm{xc}}}{\delta
n(x)}~.
\end{equation}%
So far this is exact. However, to make further progress we note that
although exchange correlations are intrinsically non-local, for a thermal
system we can assume that entanglement effects are appreciable only over
short distances. Therefore it might not be unreasonable to approximate $F_{%
\textrm{xc}}$ by a sum over independent volume elements. Accordingly, we adopt
the so-called local density approximation, 
\begin{equation}
F_{\textrm{xc}}(\beta ;n]\approx F_{\textrm{xc}}^{LDA}(\beta ;n]=\int d^{3}x\,f_{%
\textrm{xc}}(n(x))n(x)~,
\end{equation}%
where the function $f_{\textrm{xc}}(n)$ is assumed known: it is the exchange
correlation free energy per particle for a uniform electron gas with density 
$n$. The corresponding potential 
\begin{equation}
v_{\textrm{xc}}(x)=\left. \frac{d}{du}\left( f_{\textrm{xc}}(u)u\right)
\right\vert _{u=n(x)}
\end{equation}%
is therefore also known.

To find the optimal density $n^{\ast }(x)$ that solves the variational
equation (\ref{eDFT e}) we can use the same trick introduced by Kohn and
Sham. They noticed that their variational equation for the ground state ---
the analogue of our eq.(\ref{eDFT e}) --- is exactly of the form one obtains
for a gas of non-interacting and uncorrelated particles moving in an
effective single-particle potential. This leads us to rewrite (\ref{eDFT e})
as 
\begin{equation}
\left[ \frac{\delta F_{0}}{\delta n(x)}+v_{\textrm{eff}}(x)\right] _{n^{\ast
}(x)}=0~,
\end{equation}%
where 
\begin{equation}
v_{\textrm{eff}}(x)=v(x)+\int d^{3}x^{\prime }\frac{e^{2}n(x^{\prime })}{%
|x-x^{\prime }|}+v_{\textrm{xc}}(x)~.  \label{veff}
\end{equation}%
Thus, the problem of $N$ interacting particles has been translated into the
problem of a single particle moving in an density-dependent effective
potential created by all the other particles. This shows that we can adopt
the same iterative procedure followed with the Hartree self-consistent
potential. If $n^{(j)}(x)$ is the density at the $j^{{th}}$ iteration,
use (\ref{veff}) to construct the potential $v_{\textrm{eff}}^{(j)}(x)$, and
solve the single-particle equation, 
\begin{equation}
\left[ -\frac{1}{2}\nabla ^{2}+v_{\textrm{eff}}^{(j)}(x)\right] \psi
_{k}^{(j)}(x)=\varepsilon _{k}^{(j)}\psi _{k}^{(j)}(x)~.
\end{equation}%
Then construct the density $n^{(j+1)}(x)$ for the next iteration as the
thermal average, 
\begin{equation}
n^{(j+1)}(x)=\sum_{k=1}^{k_{{max}}}\frac{|\psi _{k}^{(j)}(x)|^{2}}{%
1+\exp [\beta (\varepsilon _{k}^{(j)}-\mu )]}
\end{equation}%
where the cutoff $k_{{max}}$ is such the occupation of orbitals with $%
k>k_{{max}}$ can be neglected and $\mu $ is found by imposing $\int
d^{3}x\,n(x)=N$. The process is repeated until convergence to the optimal $%
n^{\ast }$ is achieved.

Just as in the standard Kohn-Sham model neither the single particle
potential $v_{\textrm{eff}}(x)$, nor the wave functions $\psi _{k}$ and
energies $\varepsilon _{k}$ are to be given any real physical
interpretation. They are auxiliary quantities whose only purpose is the
calculation of the physical density $n^{\ast }(x)$.

\section{Conclusion}


We have produced a \emph{reconstruction} of DFT that makes explicit how DFT
fits within an ongoing research program that places the concepts of entropy
and information at the very foundation for all of physics (see \emph{e.g., }%
\cite{Caticha 2023}). This includes statistical mechanics \cite{Brillouin
1952}-\cite{Jaynes 1979}, quantum mechanics \cite{Caticha 2019}\cite{Caticha
2021a}, and as we have shown in this work, also the main techniques to study
structure --- variational principles including mean field methods and DFT.

We extended the use of entropy as a systematic method to generate optimal
approximations from the classical to the quantum domain. This allowed an
entropic reconstruction of quantum DFT. This process involves a family of
trial density operators parametrized by the particle density. The optimal
density operator is found by maximizing the quantum entropy relative to the
exact canonical density operator. This approach reproduces the variational
principle of DFT and allows a proof of the Hohenberg-Kohn theorem at finite
temperature that is simpler in that it evades some of the subtleties of the
ground state formalism. Our formalism differs from previous approaches in
that (i) the central role of entropy is explicit, and (ii) we remain with
the canonical ensemble formalism.

\textbf{Acknowledgments:} We are thankful to Oleg Lunin, Carlo Cafaro,
Herbert Fotso, and Daniel Robins, for their insightful comments.

\textbf{Conflicts of Interest:} The authors declare no conflict of interest.

\end{document}